\documentclass[twocolumn]{aastex62}

\received{28 September 2018}
\revised{20 November 2018}
\accepted{30 November 2018}
\shorttitle{Limitations of Gyrochronology}
\shortauthors{Metcalfe \& Egeland}

\begin{document}

\title{UNDERSTANDING THE LIMITATIONS OF GYROCHRONOLOGY FOR OLD FIELD STARS}

\author[0000-0003-4034-0416]{Travis S.\ Metcalfe}
\affiliation{Space Science Institute, 4750 Walnut St., Suite 205, Boulder, CO 80301, USA}
\affiliation{Max-Planck-Institut f\"ur Sonnensystemforschung, Justus-von-Liebig-Weg 3, 37077, G\"ottingen, Germany}

\author[0000-0002-4996-0753]{Ricky Egeland}
\affiliation{High Altitude Observatory, National Center for Atmospheric Research, P.O.\ 
Box 3000, Boulder, CO 80307, USA}

\begin{abstract}
Nearly half a century has passed since the initial indications that stellar rotation slows 
while chromospheric activity weakens with a power-law dependence on age, the so-called 
Skumanich relations. Subsequent characterization of the mass-dependence of this behavior 
up to the age of the Sun led to the advent of gyrochronology, which uses the rotation rate 
of a star to infer its age from an empirical calibration. The efficacy of the method 
relies on predictable angular momentum loss from a stellar wind entrained in the 
large-scale magnetic field produced by global dynamo action. Recent observational evidence 
suggests that the global dynamo begins to shut down near the middle of a star's 
main-sequence lifetime, leading to a disruption in the production of large-scale magnetic 
field, a dramatic reduction in angular momentum loss, and a breakdown of gyrochronology 
relations. For solar-type stars this transition appears to occur near the age of the Sun, 
when rotation becomes too slow to imprint Coriolis forces on the global convective 
patterns, reducing the shear induced by differential rotation, and disrupting the 
large-scale dynamo. We use data from \cite{Barnes2007} to reveal the signature of this 
transition in the observations that were originally used to validate gyrochronology. 
We propose that chromospheric activity may ultimately provide a more reliable age 
indicator for older stars, and we suggest that asteroseismology can be used to help 
calibrate activity-age relations for field stars beyond the middle of their main-sequence 
lifetimes.
\end{abstract}

\keywords{stars: activity---stars: evolution---stars: magnetic field---stars: rotation---stars: solar-type}

%%%%%%%%%%%%%%%%%%%%%%%%%%%%%%%%%%%%%%%%%%%%%%%%%%%%%%%%%%%%%%%%%%%%%%%%%% 
\section{Background}\label{sec1}

Stars are born with a range of initial rotation rates and magnetic field strengths, and 
beyond the saturated regime the two properties are intricately linked for as long as a 
global dynamo continues to operate. The large-scale magnetic field gradually slows the 
rotation over time \citep[e.g., see][]{Reville2015, Garraffo2016}. Through a process 
known as magnetic braking, charged particles in the stellar wind follow 
the magnetic field lines out to the Alv\'en radius, shedding angular momentum in the 
process. In turn, non-uniform rotation modifies the morphology of the magnetic field 
\citep[e.g., see][]{Brown2010}. Solar-like differential rotation, with a faster equator 
and slower poles, is a natural consequence of convection in the presence of substantial 
Coriolis forces \citep{Miesch2005}. The resulting shear wraps up the large-scale poloidal 
field into a toroidal configuration that ultimately leads to the emergence of active 
regions on the surface. Through these basic physical processes, stellar rotation and 
magnetism diminish together over time, each feeding off the other. The mutual feedback 
can continue as long as rotation and magnetism are coupled through a global dynamo\footnote{
By ``global dynamo'' we mean the mechanism that generates large-scale magnetic field, 
as opposed to a ``local dynamo'' which may generate field on smaller scales.}.

Nearly half a century ago, \cite{Skumanich1972} planted the observational seeds of this 
consensus view of magnetic stellar evolution. Both the theoretical foundations and the 
constraints from young clusters improved steadily over the intervening decades 
\citep[e.g., see][and references therein]{Soderblom1993}. But the Sun remained the oldest 
calibrator, so the empirical relations were largely untested beyond stellar middle-age. 
\cite{Barnes2007} put forward a more quantitative formulation of the rotation-age relation 
(so-called {\it gyrochronology}), establishing the mass-dependence of stellar spin-down 
from observations of young clusters and using the Sun to determine the age-dependence. 
Given only the B$-$V color and rotation period ($P_{\rm rot}$) of a star, gyrochronology 
yielded an empirical stellar age with a precision of 15--20\%. \cite{Barnes2010} revised 
this formulation to account for varying initial conditions ($P_0$, important in young 
clusters), and to map the mass-dependence onto a convective turnover time ($\tau_c$) 
derived from the stellar models of \cite{BarnesKim2010}. This approach more faithfully 
reproduced the distribution of rotation periods in young clusters, while yielding ages 
compatible with \cite{Barnes2007} for more evolved stars.

Observations from the {\it Kepler} mission provided the first tests of gyrochronology for 
older clusters and for field stars beyond the age of the Sun. \cite{Meibom2011} found good 
agreement with expectations for the 1\,Gyr cluster NGC\,6811, and \cite{Meibom2015} 
extended this success to 2.5\,Gyr with observations of the cluster NGC\,6819. The first 
indications of unexpected behavior were uncovered by \cite{Angus2015}, who found that no 
single gyrochronology relation could simultaneously explain the cluster data and the 
asteroseismic ages for old {\it Kepler} field stars with measured rotation periods. 
\cite{vanSaders2016} confirmed anomalously fast rotation among the best characterized {\it 
Kepler} asteroseismic targets, and proposed a model that could explain the observations 
with significantly weakened magnetic braking beyond the middle of a star's main-sequence 
lifetime. \cite{Metcalfe2016} found the magnetic counterpart of this rotational transition 
in chromospheric activity measurements of the {\it Kepler} targets, showing 
empirically that the activity level continues to decrease while the rotation rate remains 
almost constant. They suggested that the transition
might be triggered by a change in the character of differential rotation that was expected 
from global convection simulations \citep{Gastine2014, Brun2017}. \cite{Metcalfe2017} 
identified a coincident shift in stellar cycle properties, with the cycle period 
growing longer and the amplitude becoming weaker at nearly constant rotation.

These developments suggest a revised picture of the late stages of magnetic stellar 
evolution, in which the disruption of differential rotation in the absence of substantial 
Coriolis forces leads to a gradual decrease in the production of large-scale magnetic 
fields by the global dynamo. The consequence of this transition is a decoupling of 
rotation and magnetism near middle-age, such that magnetic braking can no longer shed 
angular momentum efficiently and rotation remains almost constant until the subgiant 
phase. This scenario would also explain the long-period edge found by 
\cite{McQuillan2014} in the distribution of rotation periods with B$-$V color for 34,000 
stars in the {\it Kepler} field, where significantly longer rotation periods are expected 
from gyrochronology but not observed for solar-type stars \citep{vanSaders2018}.

In an effort to address any skepticism about the existence of this transition, we 
identify its manifestation among the most evolved dwarfs in the Mount Wilson sample that 
were originally used to validate gyrochronology (Section~\ref{sec2}). We then search 
within the {\it Kepler} asteroseismic sample to identify analogs of the Mount Wilson 
stars that show the largest inconsistencies between gyrochronology and chromospheric ages 
(Section~\ref{sec3}), allowing us to characterize more precisely the decoupling of 
rotation and magnetism. In Section~\ref{sec4}, we discuss future observations of these 
Mount Wilson stars with the Transiting Exoplanet Survey Satellite (TESS), and we predict 
that their asteroseismic ages will significantly exceed those expected from 
gyrochronology. We conclude in Section~\ref{sec5} with a discussion of the potential for 
reliable chromospheric ages of older stars, using asteroseismology to recalibrate the 
activity-age relation.

%%%%%%%%%%%%%%%%%%%%%%%%%%%%%%%%%%%%%%%%%%%%%%%%%%%%%%%%%%%%%%%%%%%%%%%%%% 
\section{Gyrochronology Sample}\label{sec2}

After calibrating gyrochronology with young clusters and the Sun, \citet[][hereafter 
B07]{Barnes2007} attempted to validate the method using a sample of bright field stars 
observed for decades by the Mount Wilson HK project \citep{Wilson1968}. For the subset of 
71 stars that were not known to be significantly evolved, B07 compiled B$-$V colors, 
rotation periods $P_{\rm rot}$, and mean chromospheric activity levels $\log \left<R'_{\rm 
HK}\right>$ from the literature \citep{Noyes1984, Donahue1996, Baliunas1996}. B07 then 
used the mean activity levels to calculate chromospheric ages from the activity-age 
relation of \cite{Donahue1998}, which could be compared to the ages and uncertainties from 
gyrochronology (see Table~3 of B07). Although \cite{Donahue1998} did not provide a method 
for assessing uncertainties on the calculated ages, B07 noted that discrepancies in the 
age estimates for members of wide binaries and triple systems suggested a mean fractional 
error of 46\%, about 3 times the typical age uncertainties from gyrochronology. In 
summarizing the comparison between the two age estimates, B07 noted in the abstract: {\it 
``Gyro ages for the Mount Wilson stars are shown to be in good agreement with 
chromospheric ages for all but the bluest stars''}. Below, we use the data from Table~3 of 
B07 to reassess this comparison.

  % FIGURE 1 --------------------------------------------------------------- 
  \begin{figure*} \centerline{\includegraphics[angle=270,width=5.0in]{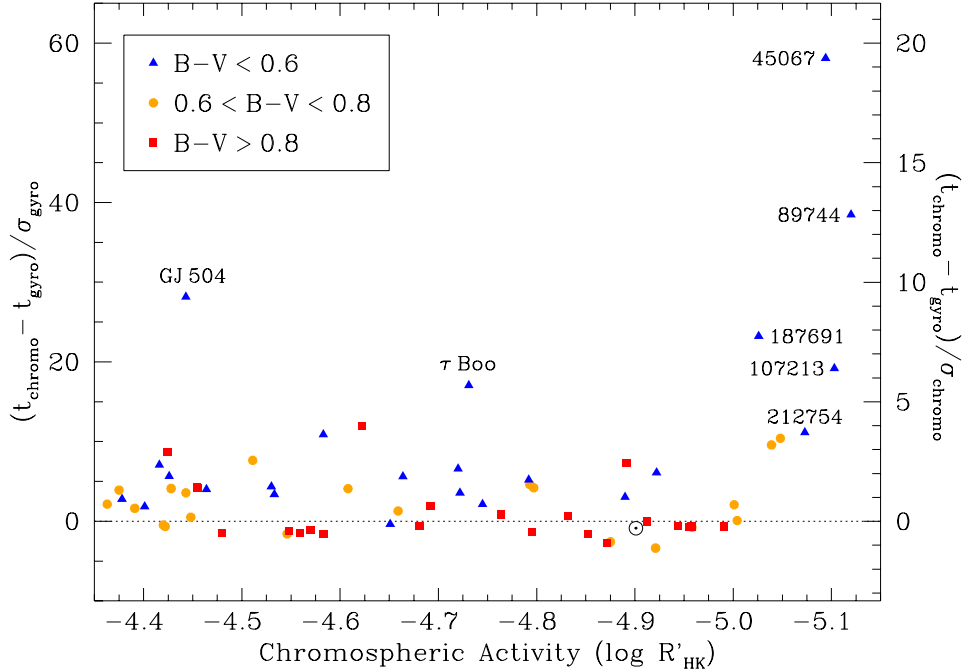}} 
  \caption{Difference between the chromospheric age and the gyro age, in units of the 
  uncertainties ($\sigma_{\rm gyro}$ and $\sigma_{\rm chromo}$), for Mount Wilson stars at 
  various activity levels. Data are taken directly from Table~3 of \cite{Barnes2007}. 
  Colored symbols indicate hotter stars with 
  B$-$V\,$<$\,0.6 (blue triangles), solar-type stars with 0.6\,$<$\,B$-$V$\,<\,$0.8 (yellow 
  circles), and cooler stars with B$-$V\,$>$\,0.8 (red squares), while the Sun is shown 
  with its usual symbol ($\odot$). With the exception of two Jovian exoplanet host stars 
  (GJ\,504 and $\tau$\,Boo), the two age estimates generally agree within 
  $\sim$3\,$\sigma_{\rm chromo}$ ($\sim$9\,$\sigma_{\rm gyro}$) until the lowest activity 
  levels ($\log R'_{\rm HK}$\,$<$\,$-$5) where rotation appears to decouple from activity 
  (see text for details). Most of these F- and G-type stars are classified as ``Flat'' or 
  ``Long'' by \cite{Baliunas1995} from 25 years of Ca~HK observations, suggesting that 
  their global dynamos may already be shutting down, eliminating the large-scale fields 
  that dominate magnetic braking. See Table~\ref{tab1} for numerical values of the 
  age discrepancy for the outliers at low activity levels.\\ \label{fig1}}
  \end{figure*} 
  %-------------------------------------------------------------------------

In Figure~\ref{fig1}, we show the difference between the chromospheric and gyro ages, in 
units of the age uncertainty, plotted against the mean chromospheric activity level for 70 
of the Mount Wilson stars tabulated in B07 (we omit only the M dwarf HD\,95735, 
which has a spurious chromospheric age of 20\,Gyr). On the 
left axis the inconsistency between the age estimates is shown in units of the tabulated 
uncertainty on the gyro age, $\sigma_{\rm gyro}$. On the right axis, the values are scaled 
to reflect the larger uncertainty on the chromospheric age, $\sigma_{\rm chromo}$. 
Colored symbols separate the sample into hotter stars with B$-$V\,$<$\,0.6 (blue 
triangles), solar-type stars with 0.6\,$<$\,B$-$V$\,<\,$0.8 (yellow circles), and cooler 
stars with B$-$V\,$>$\,0.8 (red squares). The solar values from B07 are indicated with the 
$\odot$ symbol. As noted by B07: {\it ``apart from a slight tendency toward shorter gyro 
ages... there is general agreement between the chromospheric and gyro ages for this 
sample''}. With few exceptions\footnote{Exceptions include the two Jovian exoplanet host 
stars GJ\,504 \citep{Kuzuhara2013} and $\tau$\,Boo \citep{Walker2008}, where the stellar 
rotation, activity, or both could plausibly be affected by interactions with the planet.}, 
the two age estimates tend to agree between activity levels of $-$4.3 and $-$5.0, in a 
band of uncertainty that stretches from $-$1 to $+$3 $\sigma_{\rm chromo}$ ($-$3 to $+$9 
$\sigma_{\rm gyro}$).

  % TABLE 1 ---------------------------------------------------------------- 
  \floattable
  \begin{deluxetable}{lccccc}
  \tablecaption{Mount Wilson Stars with Suspect Gyrochronology Ages\label{tab1}}
  \tablehead{\colhead{\hspace*{1.5in}} & \colhead{~~HD\,45067~~~} & \colhead{~~HD\,89744~~~} & \colhead{~~HD\,107213~~} & \colhead{~~HD\,187691~~} & \colhead{~~HD\,212754~~}} 
  \startdata
  B$-$V\ \dotfill                    & 0.56           & 0.54          & 0.50          & 0.55          & 0.52          \\
  $P_{\rm rot}$~[d]\ \dotfill        & 8              & 9             & 9             & 10            & 12            \\
  $\log \left<R'_{\rm HK}\right>$\ \dotfill& $-$5.094 & $-$5.120      & $-$5.103      & $-$5.026      & $-$5.073      \\
  $t_{\rm gyro}$~[Gyr]\ \dotfill     & $0.76\pm0.12$  & $1.11\pm0.19$ & $1.63\pm0.33$ & $1.25\pm0.21$ & $2.30\pm0.44$ \\
  $t_{\rm chromo}$~[Gyr]\ \dotfill   & 7.733          & 8.421         & 7.966         & 6.128         & 7.207         \\
  $T_{\rm eff}$~[K]\ \dotfill        & 5973           & 6149          & 6249          & 6059          & 6210          \\
  $\log g$\ \dotfill                 & 3.88           & 3.94          & 4.13          & 4.06          & 3.88          \\
  \relax [Fe/H]\ \dotfill            & $-$0.19        & $+$0.08       & $+$0.16       & $+$0.04       & $-$0.05       \\
  $L/L_\odot$\ \dotfill              & $4.18\pm0.02$  & $6.38\pm0.02$ & $5.31\pm0.02$ & $2.95\pm0.01$ & $6.72\pm0.05$ \\
  MWO $P_{\rm cyc}$~[yr]\ \dotfill   & Flat           & Flat          & Long          & 5.4~(Fair)    & Long          \\
  \enddata
  \tablerefs{\cite{Barnes2007, Boeche2016, Gaia2018, Baliunas1995}}
  \end{deluxetable}
  %-------------------------------------------------------------------------

Given the relative precision of the two methods, we can assume that most of the scatter 
is due to the chromospheric age uncertainties. Although the median age inconsistency is 
indeed slightly higher for the blue stars (B$-$V\,$<$\,0.6), the most significant 
outliers\footnote{Note that the identification of these stars as the largest outliers 
does not depend on the B07 formulation of gyrochronology. Updating all of the gyro ages 
to those produced by the \citet[][hereafter B10]{Barnes2010} formulation (and adopting 
the uncertainties from B07, since there is no prescription for calculating uncertainties 
in B10) yields the same conclusion.} are found at the lowest activity levels ($\log 
R'_{\rm HK}$\,$<$\,$-$5). These stars are labeled with their HD numbers in 
Figure~\ref{fig1}, and their properties are listed in Table~\ref{tab1}. There is reason 
to be skeptical of the chromospheric ages for these Mount Wilson stars. As pointed out by 
B07, the chromospheric ages for these stars exceed the main-sequence lifetime of a 
typical F-type star. This argument was used by B07 as justification for giving preference 
to the gyro ages. However, even if the chromospheric ages are substantially 
overestimated, the gyro ages are not necessarily correct.

Setting aside questions about the absolute reliability of chromospheric ages,
the activity levels certainly suggest that these stars might be 
significantly evolved \citep[e.g., see][]{Wright2004}. In the bottom half of 
Table~\ref{tab1}, we provide three additional lines of evidence that support this general 
conclusion. First, we list spectroscopic parameters ($T_{\rm eff}, \log g$, [Fe/H]) from 
\cite{Boeche2016}. \cite{Barnes2016} argue that gyrochronology has only been calibrated 
for dwarf stars near solar-metallicity. They classify as subgiants any star with $\log 
g<4.2$, and remove from consideration metal-poor stars. By these definitions, all of 
these Mount Wilson stars would be considered subgiants, and one of them (HD\,45067) 
should be discarded by virtue of its low metallicity. Second, the luminosities shown in 
Table~\ref{tab1} \citep{Gaia2018} are substantially above the main-sequence luminosity for 
F-type stars (which is typically less than 2~$L_\odot$). Third, after 25 years of 
monitoring their chromospheric activity, \cite{Baliunas1995} classified most of these 
stars as ``Flat'' (i.e.\ showing constant activity with fractional variations less than 
1.5\%) or ``Long'' (i.e.\ showing potential variability on a timescale longer than 25 
years), suggesting that their global dynamos may have already started to shut down 
\citep{Metcalfe2017}. The one exception is a possible 5.4~year cycle in HD\,187691, which 
was assigned a false-alarm probability (FAP) grade of ``Fair'' by \cite{Baliunas1995}. 
Note that the two solar-type stars near HD\,212754 in Figure~\ref{fig1} (HD\,178428 and 
HD\,143761) are also classified as ``Flat'' and ``Long''. Although there may be
substantial problems with chromospheric ages at these low activity levels \citep[e.g., 
see][]{Mamajek2008}, the corroborating evidence in Table~\ref{tab1} of significant 
evolution suggests that gyro ages may also suffer from systematic errors in this regime.

In the context of our revised picture of magnetic stellar evolution (Section~\ref{sec1}), 
how can we understand this inconsistency between the chromospheric and gyro ages for the 
Mount Wilson stars in Table~\ref{tab1}? As discussed by \cite{vanSaders2016}, the 
shutdown of magnetic braking appears to occur at a critical value of the Rossby number 
(Ro\,$\equiv P_{\rm rot}/\tau_c$), the ratio of the rotation period to the convective 
turnover time. Hotter stars have shallower convection zones with shorter turnover times, 
so they reach the critical Rossby number at earlier absolute ages while their rotation 
periods are still relatively short. The result of this transition is a decoupling of 
rotation and magnetism, with the rotation period remaining almost constant while the 
chromospheric activity continues to decrease\footnote{If magnetic energy driven by 
rotation on large scales is replaced with mechanical energy driven by convection on small 
scales \citep{BohmVitense2007}, then the change in magnetic morphology that dramatically 
reduces angular momentum loss need not change the chromospheric activity level abruptly.
Recent simulations by \cite{Garraffo2018} support this interpretation.}
with age \citep{Metcalfe2016}. As suggested by \cite{Metcalfe2017}, stellar cycles appear 
to grow longer and decrease their amplitude during this transition before disappearing 
entirely or becoming undetectable, leading to classifications of ``Long'' or ``Flat'' in 
stellar cycle surveys. With the rotation period essentially fixed, the gyro age is a 
lower limit that actually reflects the age when the star stopped spinning down in the 
middle of its main-sequence lifetime.

  % TABLE 2 ---------------------------------------------------------------- 
  \floattable
  \begin{deluxetable}{lccc|c}
  \tablecaption{Kepler Analogs of the Gyrochronology Outliers\label{tab2}}
  \tablehead{\colhead{\hspace*{1.5in}} & \colhead{~~KIC\,9139151~~~} & \colhead{~~KIC\,12009504~~} & \colhead{~~KIC\,10963065~~} & \colhead{~~KIC\,10909629~~}}
  \startdata
  B$-$V\ \dotfill                  & 0.520          & 0.556         & 0.509          & 0.540          \\
  $P_{\rm rot}$~[d]\ \dotfill      & $10.96\pm2.22$ & $9.39\pm0.68$ & $12.58\pm1.70$ & $12.37\pm1.22$ \\
  $\log R'_{\rm HK}$\ \dotfill     & $-$4.954       & $-$4.977      & $-$5.054       & $\cdots$       \\
  $t_{\rm gyro}$~[Gyr]\ \dotfill   & $1.93\pm0.37$  & $1.06\pm0.18$ & $2.82\pm0.57$  & $2.04\pm0.37$  \\
  $t_{\rm astero}$~[Gyr]\ \dotfill & $1.94\pm0.31$  & $3.44\pm0.44$ & $4.33\pm0.30$  & $4.69\pm0.56$  \\
  $t_{\rm chromo}$~[Gyr]\ \dotfill & 4.73           & 5.15          & 6.75           & $\cdots$       \\
  $T_{\rm eff}$~[K]\ \dotfill      & 6302           & 6179          & 6140           & 6265           \\
  $\log g$\ \dotfill               & 4.38           & 4.21          & 4.29           & 3.90           \\
  \relax [Fe/H]\ \dotfill          & $+$0.10        & $-$0.08       & $-$0.19        & $-$0.12        \\
  $L/L_\odot$\ \dotfill            & $1.71\pm0.01$  & $2.71\pm0.01$ & $1.93\pm0.01$  & $6.59\pm0.11$  \\
  \enddata
  \tablerefs{\cite{Garcia2014, Metcalfe2016, Creevey2017, Serenelli2017, Buchhave2015, Gaia2018}}
  \end{deluxetable}
  %-------------------------------------------------------------------------

%%%%%%%%%%%%%%%%%%%%%%%%%%%%%%%%%%%%%%%%%%%%%%%%%%%%%%%%%%%%%%%%%%%%%%%%%% 
\section{Analogs Observed by {\it Kepler}}\label{sec3}

Although asteroseismic observations do not yet exist for the Mount Wilson stars in 
Table~\ref{tab1} (see Section~\ref{sec4}), we can search for analogs of these stars within 
the sample of {\it Kepler} asteroseismic targets. There are currently 18 {\it Kepler} 
targets with detailed asteroseismic modeling (for precise ages) that also have known 
rotation periods and measured chromospheric activity \citep[see][their 
Table~1]{Metcalfe2016}. Among these stars, only three fall within the same range of 
rotation periods and B$-$V colors as the Mount Wilson stars in Table~\ref{tab1}. We list 
the properties of these analogs in the first three columns of Table~\ref{tab2}, ordered by 
activity level.

The chromospheric activity levels for these analogs span the range where the magnetic 
transition summarized in Section~\ref{sec1} is expected to occur. The critical Rossby 
number found by \cite{vanSaders2016} can also be understood as a critical activity level, 
because the two properties are strongly correlated \citep{Mamajek2008}. The critical 
activity level is around $-$4.95 \citep{Brandenburg2017}, so the gap between 
gyrochronology and other age estimates is expected to grow wider as stars continue to 
evolve to lower activity levels. We use the B$-$V color and $P_{\rm rot}$ to calculate 
gyro ages and uncertainties using the B07 formulation\footnote{Ages calculated with the 
B10 formulation agree with those from B07 within 2$\sigma_{\rm gyro}$.}, and we use $\log 
R'_{\rm HK}$ to calculate chromospheric ages following \cite{Donahue1998}. For 
KIC\,9139151, which is just reaching the critical activity level where rotation and 
magnetism are expected to decouple, the gyro age agrees with the asteroseismic age. For 
the more evolved dwarfs KIC\,12009504 and KIC\,10963065, the gap between the gyro age and 
the asteroseismic age is significant. The gyro age for these stars appears to indicate the 
point at which magnetic braking became inefficient and rotation stopped evolving 
substantially. The available data from {\it Kepler} suggests that gyrochronology is an 
unreliable age indicator for hotter stars beyond $\sim$2--3~Gyr \citep{vanSaders2016}.

The chromospheric ages for the analogs appear to be substantially overestimated at these 
low activity levels, just as with the Mount Wilson stars in Table~\ref{tab1}. The other 
indicators of evolutionary status that are listed in the first three columns of 
Table~\ref{tab2} generally fall within the range where gyrochronology has been 
calibrated. The surface gravities are all above the cut ($\log g>4.2$) suggested by 
\cite{Barnes2016}, only one of the stars (KIC\,10963065) is significantly metal-poor, and 
the Gaia luminosities are well below those of the Mount Wilson stars. Most significantly, 
there is no apparent reason to expect gyrochronology to fail for KIC\,12009504, but both 
the B07 (1.06~Gyr) and B10 (1.10~Gyr) ages are wildly inconsistent with the asteroseismic 
age \citep[modeling results vary between 3.10--4.12~Gyr,][]{SilvaAguirre2017}. If this 
star is actually too evolved for gyrochronology to be reliable, the rotation period 
should have {\it slowed} as it expanded into a subgiant, biasing the gyro age {\it older} 
and reducing the inconsistency with asteroseismology.

  % FIGURE 2 ---------------------------------------------------------------
  \begin{figure*}
  \centerline{\includegraphics[angle=270,width=5.0in]{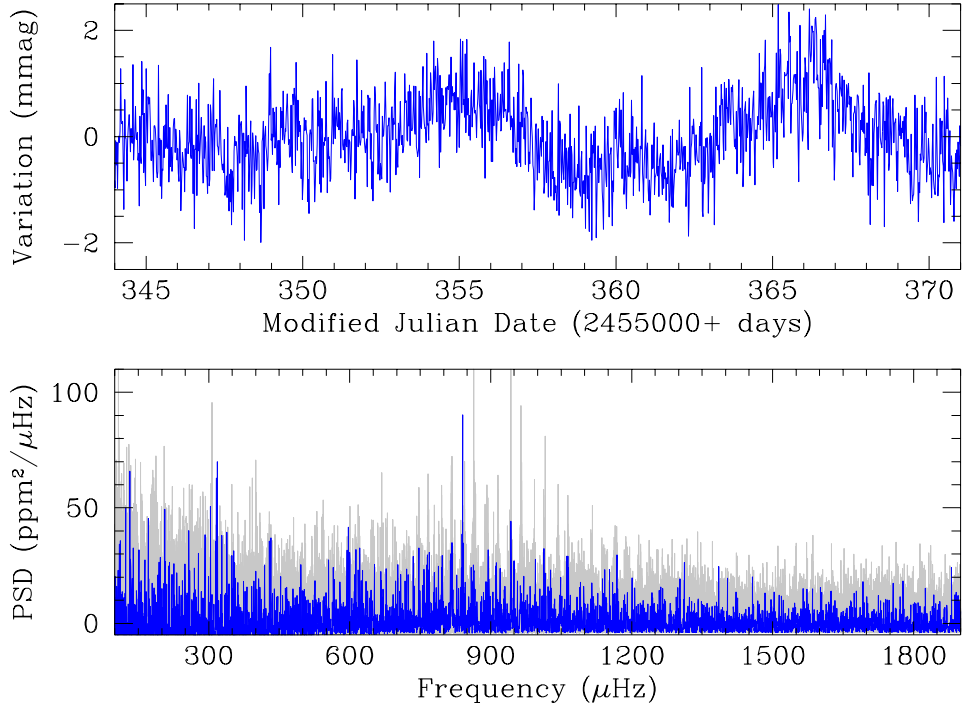}}
  \caption{Anticipated quality of TESS observations for the Mount Wilson stars listed in Table~\ref{tab1}. 
  {\bf Top:} calibrated 27-day light curve of KIC\,10909629, which has a color (B$-$V=0.54) 
  and rotation period \citep[12.37~days;][]{Garcia2014} similar to HD\,212754. With a 
  Kepler magnitude Kp=10.9, this target yields a S/N comparable to that expected from TESS 
  at V$\sim$5.9. The rotational modulation is apparent, even in this short time series. 
  {\bf Bottom:} power spectral density from 390 days of short-cadence data for KIC\,10909629 (gray) 
  and for the observations spanning 30 days (blue) that were originally used to detect solar-like 
  oscillations \citep{Chaplin2011}. The signatures of activity, granulation, and shot noise 
  have been removed using the A2Z pipeline \citep{Mathur2010}. The power excess 
  $\sim$900~$\mu$Hz is significant even in 30 days \citep{Chaplin2014}.\\ \label{fig2}}
  \end{figure*}
  %-------------------------------------------------------------------------

%%%%%%%%%%%%%%%%%%%%%%%%%%%%%%%%%%%%%%%%%%%%%%%%%%%%%%%%%%%%%%%%%%%%%%%%%% 
\section{Predictions for TESS}\label{sec4}

The TESS mission launched successfully in April 2018, and it is expected to gather 
short-cadence photometry (2-minute sampling) for all of the Mount Wilson stars listed in 
Table~\ref{tab1}. Detections of solar-like oscillations comparable to what was achieved 
by {\it Kepler} are expected in TESS targets that are $\sim$5 magnitudes brighter 
\citep{Campante2016}. The minimum dwell time on each TESS sector is 27 days, comparable 
to the 30-day time series obtained for {\it Kepler} targets during the asteroseismic 
survey that was conducted in the first year \citep{Chaplin2011}. The amplitude of 
solar-like oscillations scales with the ratio of luminosity to mass \citep{Houdek1999}, 
and detections are more likely in magnetically inactive stars \citep{Chaplin2011b}, so 
the F-type stars in Table~\ref{tab1} should yield asteroseismic data comparable to {\it 
Kepler} targets in the magnitude range Kp$\sim$10--11. The 27-day time series should also 
allow an independent check of the rotation periods determined from the Mount Wilson data.

We searched the \cite{Garcia2014} rotation catalog for the fainter {\it Kepler} 
asteroseismic targets that can be considered analogs of the Mount Wilson stars in terms 
of both stellar properties and the expected data quality from TESS. The properties of 
KIC\,10909629, the best {\it Kepler} analog of HD\,212754, are listed in the fourth 
column of Table~\ref{tab2}. KIC\,10909629 has a B$-$V color and rotation period that are 
both similar to HD\,212754, but it is 5.25 magnitudes fainter in the V band. The 
chromospheric activity level of KIC\,10909629 is not known, but the photospheric activity 
proxy\footnote{The photospheric activity proxy can be artificially low for stars when the 
rotation is viewed nearly pole-on.} \citep[$S_{\rm ph}$,][]{Mathur2014} is comparable to 
that of KIC\,10963065 which has $\log R'_{\rm HK}<-5$. The gyro age from B07 (2.04~Gyr) 
and B10 (2.24~Gyr) are both significantly younger than the asteroseismic age 
\citep[4.69~Gyr,][]{Serenelli2017}. The spectroscopic parameters and Gaia luminosity are 
similar to those of HD\,212754, and support the conclusion that KIC\,10909629 is 
substantially evolved despite its young gyro age. Again, this can be understood if the 
rotation period of KIC\,10909629 stopped evolving after $\sim$2~Gyr. Based on these 
results, we predict that asteroseismic ages from TESS for the Mount Wilson stars in 
Table~\ref{tab1} will be significantly older than expected from gyrochronology.

The anticipated quality of the TESS observations for the Mount Wilson stars is 
illustrated in Figure~\ref{fig2}. In the top panel, a 27-day segment of the long-cadence 
{\it Kepler} observations of KIC\,10909629 clearly shows rotational modulation with a 
period near 12 days and a peak-to-peak amplitude of a few milli-magnitudes. In the bottom 
panel, we show the power spectrum from 390 days of short-cadence data for KIC\,10909629 
(gray) and for the observations spanning 30 days (blue) that were originally used to 
detect solar-like oscillations in this star \citep{Chaplin2011}. In both cases, the 
signatures of activity, granulation, and shot noise have been modeled and removed using 
the A2Z pipeline \citep{Mathur2010}. While the longer data set clearly shows the series 
of evenly spaced frequencies that are characteristic of solar-like oscillations, the 
shorter time series still reveals a significant power excess $\sim$900~$\mu$Hz and a 
signature of the regular spacing. When combined with spectroscopic parameters, these 
global properties of the oscillations are sufficient to constrain the stellar age with 
$\sim$10--20\% precision \citep{Chaplin2014, Serenelli2017}.

%%%%%%%%%%%%%%%%%%%%%%%%%%%%%%%%%%%%%%%%%%%%%%%%%%%%%%%%%%%%%%%%%%%%%%%%%% 
\section{Summary and Discussion}\label{sec5}

After \cite{Skumanich1972} presented observational evidence that stellar rotation rates 
and magnetic activity levels diminish together over time, the idea of using one or both 
properties to determine the ages of stars has gradually taken hold. \cite{Barnes2007, 
Barnes2010} made the idea more quantitative by establishing the mass-dependence and 
calibrating a gyrochronology relation using young clusters and the Sun. In the absence of 
additional observations, it was natural to extrapolate these relations to stars beyond the 
middle of their main-sequence lifetimes. However, over the past few years evidence has 
emerged that something unexpected occurs in the evolution of stellar rotation and 
magnetism around middle-age, limiting the utility of gyrochronology relations. When the 
rotation period of a star becomes comparable to the global convective turnover time, 
Coriolis forces can no longer sustain the solar-like pattern of differential rotation. The 
resulting loss of shear disrupts the production of large-scale magnetic field by the 
global dynamo \citep{Metcalfe2016}. The elimination of large-scale field leads to a 
dramatic reduction in the efficiency of magnetic braking, so the stellar rotation remains 
almost constant until the subgiant phase \citep{vanSaders2016, vanSaders2018}. At the 
same time the global dynamo gradually shuts down, with the activity cycle period growing 
longer while the cycle amplitude decreases before disappearing or becoming undetectable 
\citep{Metcalfe2017}.

We have identified the signature of this transition in the observations that were 
originally used to validate gyrochronology. Using data directly from \citet[][his 
Table~3]{Barnes2007}, we demonstrate that the most significant differences between 
chromospheric ages and gyrochronology occur for the most evolved F-type dwarfs 
(Figure~\ref{fig1}). We present several independent lines of evidence to corroborate this 
interpretation, including surface gravities, Gaia luminosities, and the predominant 
absence of activity cycles (Table~\ref{tab1}). We identify analogs of these F-type stars 
among the sample of asteroseismic targets observed by {\it Kepler}, and we show that the 
asteroseismic ages agree with gyrochronology until a critical activity level ($\log 
R'_{\rm HK}=-4.95$) beyond which the two estimates diverge (Table~\ref{tab2}). Finally, 
we use observations of a fainter analog from the {\it Kepler} sample to predict the 
quality of observations anticipated for these targets from the TESS mission, showing that 
rotation periods and solar-like oscillations should both be detectable 
(Figure~\ref{fig2}). Considering our revised picture of the late stages of magnetic 
stellar evolution, we predict that these future observations will demonstrate that 
gyrochronology is unreliable for stars beyond the middle of their main-sequence 
lifetimes.

There are two key updates to the scenario for magnetic evolution outlined in this paper 
compared to that proposed by \cite{Metcalfe2016}. First, it is now clear that the Rossby 
number from global convection simulations and that obtained from asteroseismic models 
that use a mixing-length prescription are not directly comparable \citep{Brun2017}. Based 
on solar determinations of the Rossby number from both methods, we now believe that the 
transition near Ro$\sim$2 identified by \cite{vanSaders2016} corresponds to a change in 
the character of differential rotation seen near Ro$\sim$1 in convection simulations. In 
the context of \cite{Metcalfe2016}, this implies that stellar evolution across the 
Vaughan-Preston gap entirely precedes the magnetic transition, which occurs at a 
substantially lower activity level. Second, there is now observational evidence of 
possible anti-solar differential rotation in some stars that show higher than expected 
activity for their rotation rates \citep{Brandenburg2018}. This suggests that when stars 
reach the critical Rossby number, the differential rotation might flip from solar-like to 
anti-solar (slow equator, fast poles). Conservation of angular momentum would require the 
shear to increase, leading to an enhancement of activity in the slowly rotating regime 
\citep{Karak2015}. Future observations will determine whether this phenomenon represents 
a temporary phase, or an alternate pathway that coexists with the shutdown of the global 
dynamo.

While it may be disappointing that rotation is less useful as a diagnostic of age beyond 
the middle of stellar main-sequence lifetimes, the other Skumanich relation 
(activity-age) may not be similarly disrupted. Observations of chromospheric activity in 
a large sample of solar analogs suggest that, unlike rotation, the evolution of activity 
appears to be continuous across the magnetic transition \citep{LorenzoOliveira2018}. 
Although the ages adopted for their analysis were derived from isochrones, the TESS 
mission is poised to provide reliable asteroseismic ages for bright stars down to 
V$\sim$7 all around the sky. When combined with existing archives of chromospheric 
activity, it is possible that asteroseismic ages can be used to recalibrate the 
activity-age relation for older solar-type stars. Given the difficulty of obtaining time 
series measurements of the diminishing rotational modulation in such stars, and 
considering that minimal chromospheric variability makes one spectroscopic measurement 
more likely to be representative of the mean activity level, chromospheric activity might 
ultimately provide a more reliable age indicator for stars beyond middle-age.

%%%%%%%%%%%%%%%%%%%%%%%%%%%%%%%%%%%%%%%%%%%%%%%%%%%%%%%%%%%%%%%%%%%%%%%%%% 

\acknowledgements {\sl The authors would like to thank Steven Blau, Axel Brandenburg, Hannah
Schunker, Andrew Skumanich, David Soderblom, and Jennifer van~Saders for helpful exchanges,
as well as Rafael Garc{\'\i}a, Savita Mathur, and Warrick Ball for assistance with
Figure~\ref{fig2}. This research made use of NASA's Astrophysics Data System, as well as
the SIMBAD database and the VizieR catalog access tool at CDS in Strasbourg, France. This
work benefitted from discussions within the international team ``The Solar and Stellar
Wind Connection: Heating processes and angular momentum loss'' at the International Space
Science Institute (ISSI). Support was provided by a Visiting Fellowship at the Max Planck
Institute for Solar System Research, and by the Nonprofit Adopt a Star program
(\href{https://adoptastar.org}{adoptastar.org}) administered by White Dwarf Research
Corporation. R.~E. is supported by an NCAR Advanced Study Program Postdoctoral
Fellowship. The National Center for Atmospheric Research is sponsored by the U.S.~National
Science Foundation.}

%%%%%%%%%%%%%%%%%%%%%%%%%%%%%%%%%%%%%%%%%%%%%%%%%%%%%%%%%%%%%%%%%%%%%%%%%% 

\end{document}